# Determination of coronal temperatures from electron density profiles

J.F. Lemaire
BISA, Brussels  (jfl@oma.be)

## Abstract

The most popular method for determining coronal temperatures is the scale-height-method (*shm*). It is based on electron density profiles inferred from White Light (WL) brightness measurements of the corona during solar eclipses. This method has been applied to several published coronal electron density models. The calculated temperature distributions reach a maximum at r > 1.3 $R_S$, and therefore do not satisfy one of the conditions for applying the *shm* method. Another method is the hydrostatic equilibrium method (*hst*), which enables coronal temperature distributions to be determined, providing solutions to the hydrostatic equilibrium equation. The temperature maximas using the *hst* method are almost equal to those obtained using the *shm* method, but the temperature peak is always at significantly lower altitude when the *hst*-method is used than when the *shm*-method is used.  A third and more recently developed method, *dyn*, can be used for the same published electron density profiles. The temperature distributions obtained using the *dyn* method are regular solutions of the hydrodynamic equations. They depend on the expansion velocity of the coronal plasma considered as a free input parameter in the calculations. The larger the solar wind expansion velocity at 1AU, the larger the new temperature maximum that develops in the range of altitudes (about 3 $R_S$) where the outward acceleration rate of the coronal plasma is greatest. At the base of the solar corona, where the coronal bulk velocity is small (subsonic), the *dyn*  and *hst* methods give similar temperature values. More significant differences are found at higher altitudes where the expansion velocity approaches and exceeds the velocity of sound. This paper discusses the effects of three factors on calculated *dyn* temperature distributions: (i) super-radial expansion flux tubes; (ii) alpha particle concentration; and (iii) different ratios of ion temperature over electron temperature. The *dyn* method is a generalization of the *hst* method, where the electron temperature distribution tends to zero at infinite radial distances. The *dyn* method discussed here is a useful diagnostic tool for determining from WL coronal observations the range of radial distances where the coronal heating rate is at its maximum.

<u>Keywords</u>: solar physics; solar corona; electron temperature distributions; methods for the determination of solar, stellar and planetary atmospheres;  super-radial expansion flow; positions of maximum heating rate in solar corona.

## 1. Introduction

For more than half a century, coronal white light (WL) brightness and polarization measurements from solar eclipses observations have been used to determine radial profiles of $n_e(r)$, the electron density in the solar corona. Values of H, the density scale height, have been derived from the gradient of the electron density. Since H is proportional to the temperature, coronal temperatures have been determined at different heliographic latitudes and for all sorts of solar activity conditions.

The scale-height-method (*shm*) has been commonly used to evaluate coronal temperatures from eclipse WL observations. The method was reviewed in a seminal article by van de Hulst [28]: it postulates that $n_e(r)$ is a solution of the hydrostatic equation and that the coronal plasma is isothermal and homogeneous. The latter assumptions must be satisfied for the *shm* method to be applicable. This implies that $T_e(r)$ is independent of the radial distance and that the coronal electron density distribution is an exponential function of r.

Additional simplifying assumptions are required to invert brightness observations and to obtain empirical distributions of $n_e(r)$. Spherical symmetry of the coronal plasma distributions, or *ad hoc* assumptions about 'filling factors' along lines-of-sight, have been postulated in the past[1]. It is assumed that these empirical electron density profiles are reliable, despite the high level of non-uniformity observed in the solar corona and radial density distributions reported for half a century by many solar physicists, such as Baumbach [6], van de Hulst [28, 29], Saito [27], Pottasch [26], Koutchmy [22] and Fisher and Guhathakurta [18] [20]. It is assumed that the line-of-sight average density scale heights they provide are satisfactory approximations characterizing average distributions of plasma in the solar corona.

As shown in the next section, <H(r)> the average line-of-sight distributions of H can be used to calculate radial profiles for $T_e(r)$ and $<T_e(r)>$. Figure 2 shows *shm* distributions of $T_e(r)$ for the electron density profiles illustrated in Figure 1.

The *shm* temperature distributions have been obtained for the equatorial and polar regions of the solar corona, as well as for coronal holes such as those studied by Munro and Jackson [24]. These *shm* temperature profiles exhibit a maximum beyond 1.3 $R_S$.

As shown in the third section, the solar corona is not in hydrostatic equilibrium when $T_e(r)$ is determined by the *shm* method. The reasons for this contradiction are that observed coronal density profiles do not decrease exponentially as a function of r, and the solar corona is not isothermal, as required for *shm* method to be applicable.

In section four, an alternative method for determining the distribution of coronal temperatures is outlined. This method is applicable when the corona is really in hydrostatic equilibrium and its density distribution is not decreasing exponentially. The temperature profiles obtained using this method will be labelled (*hst*). They differ from those obtained using the *shm* method. Although both methods give comparable values for the maximum temperature (1.1-1.5 MK), the altitudes at which the temperature peak is reached in the inner corona are significantly different. This also implies that the altitudes where the coronal heating rate is expected to be at its maximum depends on the methods employed: *shm* or *hst*.

In section five, a third method is presented for calculating coronal temperatures in the more general situations when the corona is not in hydrostatic equilibrium, but is expanding, as modeled by Parker [25] [2]. This more general method is the hydrodynamic or dynamic equilibrium method (*dyn*).

---

[1] Highest resolution coronal observations currently indicate that the coronal plasma consists mainly of thin filamentary structures. To some extent, this limits the validity of the spherical symmetry assumption. It is expected that averaging over all the filaments along lines-of-sight does not significantly affect the mean density gradients and mean scale height profiles.

In the final section, the calculated temperature distributions obtained using all these methods (*shm*, *hst* and *dyn*) are compared with each other for different solar wind expansion velocities, different geometries of funnel-shaped flux tubes, and different ratios of ion vs. electron temperatures.

## 2. The scale-height method (*shm*)

In atmospheres that are in isothermal hydrostatic equilibrium, the particle number density is distributed according to the well-known barometric formula (i.e., by an exponential function of the altitude h):

$$n(h) / n(h_o) = \exp[-(h - h_o) / H] \qquad (1)$$

In this formula, $n(h_o)$ is the density at a reference altitude, $h_o$. This equation has been applied for many decades in studies of planetary atmospheres. The density scale height, H, is defined by:

$$H = (- d \ln n / dh)^{-1} = k T / \mu\, m_H\, g \qquad (2)$$

where k is the Boltzmann constant; $m_H$ is the mass of hydrogen atom; and $\mu$ is the mean molecular mass of the gas.

H is considered as a key parameter characterizing the rate of decrease in the density of planetary and stellar atmospheres. When the density decreases slowly with altitude, H is large and T, the atmospheric temperature, is also large. This is precisely the case for the solar corona where $\mu \approx 0.5$ and H ~ 120,000 km, implying that the coronal temperature is T ~ 1 MK.

The altitude over which an actual density decreases by a factor *e* is given by eq (2) only when this density decreases exponentially, as in eq (1). This fact is often overlooked, however, when the actual density distribution is fitted by a sum of power laws of 1/r, as proposed by Baumbach [6]. It should also be noted that eq (1) is applicable only when g is independent of h, and when the radius of curvature of the atmospheric layers is very large compared with the H value. Unfortunately, none of these assumptions is applicable to the solar corona.

*Curved atmospheric layers.* Since the solar corona extends over a wide range of altitudes, g should not be assumed to be independent of r, but $g(r) = g_o / r^2$, where r is the normalized radial distance in units of the solar radius: $R_S = 6.96\ 10^{10}$ cm, and $g_o$ is the gravitational acceleration on the surface of the Sun ($g_o = 2.7\ 10^4$ cm/s$^2$).

Assuming that the coronal plasma is quasi-neutral, that the kinetic pressures of the electrons and ions are isotropic[3] and that their sum is given by $p = n_e k T_e + n_p k T_p + n_\alpha k T_\alpha = \nu_p n k T$, the hydrostatic equation becomes:

$$d(\nu_p n k T) / dr = -n(r) \mu m_H g(r) R_S \qquad (3)$$

---

[2] The radial expansion of the solar corona is caused by convective instability of the corona. It has been demonstrated by Lemaire [34] that Chapman's [10] conductive and hydrostatic model of the solar corona is convectively unstable. He showed that only a continuous outward expansion of the coronal plasma is capable of effectively evacuating the excess of energy deposited in the inner corona, and therefore of carrying this excess heat away into interplanetary space. Note that part of the energy deposited in the inner corona at and below the temperature peak flows downward into the transition region and chromosphere.

In a fully ionized $H^+$ plasma, the dimensionless parameter $\nu_p = 2$. If $\alpha = n_{He}/n_p$ is the relative concentration of the $He^{++}$ ions, $\nu_p = (2 + 3\alpha)/(1 + 2\alpha)$; for $\alpha = 10\%$, $\nu_p = 1.916$, provided that the alpha particles, protons and electrons all have the same temperature T. A more general expression is given in the Appendix by eq (A7) when the ion and electron temperatures are not equal.

When spherical symmetry is taken into account, and when T, $\mu$ and $\nu_p$ are again independent of r, the straightforward integration of the hydrostatic equilibrium eq (3) gives:

$$n(r) / n(r_o) = \exp[-(R_S / H_0)(1/r - 1/r_o)] \qquad (4)$$

It can be seen from eq (4) that Ln n is a linear function of $1/r$. The slope of this linear relationship determines the density scale height at the reference level $r_o$, where $g = g_o$: $H_o = \nu_p k T / \mu m_H g_0$.

This more general expression for n(r) is still an exponential function of r, whose importance has been popularized by van de Hulst [28] in his seminal 1950 article '*On the polar rays of the corona*'. It has therefore become common practice to use eq (4) to determine coronal temperatures. These are sometimes called 'hydrostatic temperatures' [31], but this could be misleading, as shown below.

*Some coronal density profiles derived from WL observations.* Figure 1 illustrates typical radial distributions of the coronal electron density determined from WL brightness and polarization observations during a series of solar eclipses. These observations have been selected for either their historical importance or their popularity. Baumbach's [6] equatorial density profile was among the first to be fitted in 1937 by a sum of power laws; it is labelled '*B*' in Figure 1 [4].

Pottasch's [26] equatorial density profile (labeled '*P*') has been selected because it was based on eclipse observations in 1952 during a solar minimum and because he pointed out that the equatorial coronal temperature had a flat maximum of 1.41 MK located between $r = 1.2 R_S$ and $2 R_S$.

Saito's [27] popular density model was developed in 1970 from a compendium of solar eclipse observations, all at heliographic latitudes. It has been widely used in the community for many years because it provided the first two-dimensional (2-D) distribution of average electron densities as a function not only of the radial distances, r, but also of the heliographic latitude $\Phi$. Saito's empirical 2-D model is based on average electron density distributions for minimum solar activity conditions. The equatorial and polar density of Saito's models are labelled '*Seq*' and '*Spol*' in Figure 1, respectively.

---

[3] There are indications from Doppler shift measurements of ion coronal emission lines that the heavy ion temperatures are larger in the horizontal direction than in vertical directions. This implies that the assumption of isotropy is questionable for ions. The electron velocity distribution functions are more closely isotropic because of their higher collision frequency in the solar corona. As a result, the electron kinetic pressure is likely to be nearly isotropic in the corona.

[4] In 1937, when Baumbach [6] published his empirical density model, the origin and properties of the corona had not yet been identified. The existence of two separate WL components was not yet known: the K-corona and the F-corona were not yet separated by polarization measurements. It was Allen [2] who, in 1947, corrected Baumbach's original density distribution for the effect of WL scattering by interplanetary dust particles (i.e., the contribution of F-corona to the total brightness).

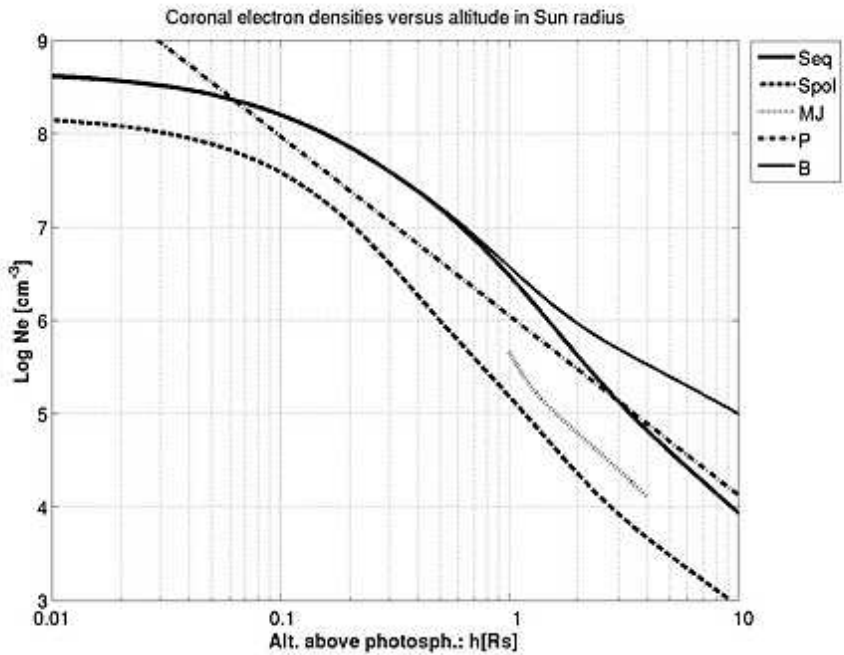

Figure 1. *Radial distribution of empirical coronal electron densities obtained from WL brightness measurements.* These empirical density models are taken from: Saito's [27] average equatorial model (*Seq*); Saito's [27] average polar model (*Spol*); Munro and Jackson's [24] coronal hole model (*MJ*); Pottasch's [26] equatorial model (*P*); and Baumbach's [6] empirical model (*B*). Log-scale is used for h, the altitude above the photosphere, as it attenuates the density variation in the inner corona. These density distributions are extended to fit solar wind densities up to 1 AU. The input parameters characterizing these models are listed in Table 1.

Munro and Jackson's [24] empirical model was among the first density distribution models for a coronal hole. It is labelled '*MJ*' in Figure 1. Note that Figure 1 confirms that coronal electron densities are smaller in coronal holes and above the poles than in the equatorial region.

Many additional empirical density profiles have been published in recent decades, such as [32], [22] and [14]. As they are rather similar to those reported above, there was no point in including them all. The main point here is to illustrate the differences resulting from various calculation methods. A larger number of published density profiles were reported by Aschwanden [5] and were shown in Figure 1.20 of his comprehensive monograph.

*Power law expansions for coronal density distributions.* As mentioned earlier, radial distributions of coronal brightness and electron densities can conveniently be fitted with sums of power laws of $1/r$. [5]

---

[5] In 1880 Schuster [33] had already hypothesized that the brightness of the solar corona was due to the scattering of sunlight by particles in the solar atmosphere. He had shown that the density of these particles decreased according to a sum of power law functions of $1/r$.

The empirical radial distribution reported by Baumbach [6] was fitted by a sum of three terms. The six constant parameters of this expansion were determined from WL coronal brightness measurements collected during eclipses between 1905 and 1927:

$$n(r) = 10^8 (2.99\, r^{-16} + 1.55\, r^{-6} + 0.036\, r^{-1.5}) \quad [\text{electrons} / \text{cm}^3] \quad (5)$$

The application of this formula should, in principle, be limited to $1.05\, R_S < r < 3\, R_S$ [6].

Separate coronal density distributions have been developed for maximum and minimum solar activity conditions, and many eclipse observers have used power laws to fit the radial distributions of electron densities in the equatorial and polar regions of the corona, in coronal streamers and in coronal holes, above quiet or active regions. For example, Brandt et al. [8] fitted four coronal distributions obtained by (i) de Jager [13] for solar minimum, (ii) Pottasch [26] for solar minimum, (iii) Allen [2] for solar maximum and (iv) van de Hulst [28] for solar maximum (see Figure 1 in [8]).

Although these empirical density profiles were extended up to $10\, R_S$ in their Figure 5, Brandt et al. [8] warned that, beyond 6-8 $R_S$, these density distributions might not be very meaningful. Further out, electron densities were extended up to 1AU by using an additional $r^{-2}$ power law term corresponding to the solar wind density at large radial distances.

A major advantage of dealing with analytical formulae such as (5) or (16) to fit observed density distributions is that $d\, Ln\, n /d(1/r)$ can then be derived analytically from n(r). This is more convenient and precise than deriving the values of grad n and H from finite difference algorithms applied to discrete values of $n(r_j)$ given at a set of positions $r_j$. Therefore, when WL coronal brightness and densities distributions are available from eclipse observations, fitting them using power laws such as (5) has clear practical advantages. These advantages have not always been appreciated, but they are clear in the following applications. A mathematical expression for the radial profiles of T(r) can then be derived from eq (6).

*Coronal electron temperature distributions determined by the scale-height-method (shm).*
In the past, many radial distributions of $n_e(r)$ and $d\, Ln\, n_e / d(1/r)$ were derived from experimental measurements of WL coronal brightness. The coronal temperature profile was then determined by:

$$T_e(r) = \mu\, m_H\, g_o\, R_S / k\, [-d\, Ln\, n_e / d(1/r)] \quad (6)$$

As mentioned earlier, the *shm* method has been often used to evaluate coronal temperatures from WL eclipse observations (see [29] [22] [32] [31]). It has been used since the 1940s to evaluate temperatures over both quiet and active regions of the Sun along coronal streamers, polar coronal rays and in coronal holes, as well as in fine-scale inter-plume structures of the corona [18] [20]. Ground-based coronagraphs, as well as WL coronagraphs on spacecraft, have been used to observe the low and high latitude regions of the solar corona and to determine coronal temperatures using the *shm*-method (see [28] [29] [7] [22] [19] [31]).

*Radial temperature distributions from the shm method.* Figure 2 shows the electron temperature distributions obtained using the *shm* method for the density distributions shown in Figure1.

---

[6] A similar mathematical expression was adopted by Allen [2] using the eclipse observations reported by Baumbach [6] but taking into account the spurious scattering of WL by interplanetary dust grains (F-corona) whose contribution can be inferred from polarization measurements.

These temperature profiles have a maximum beyond h = 0.1 $R_S$ (i.e., r > 1.1 $R_S$). Note that this altitude is well above the top of the transition region located at $h_{TR}$ = 2500 km = 0.0036 $R_S$ above the photosphere. The peak coronal temperature is larger than 1 MK in all examples shown in Figure 2.

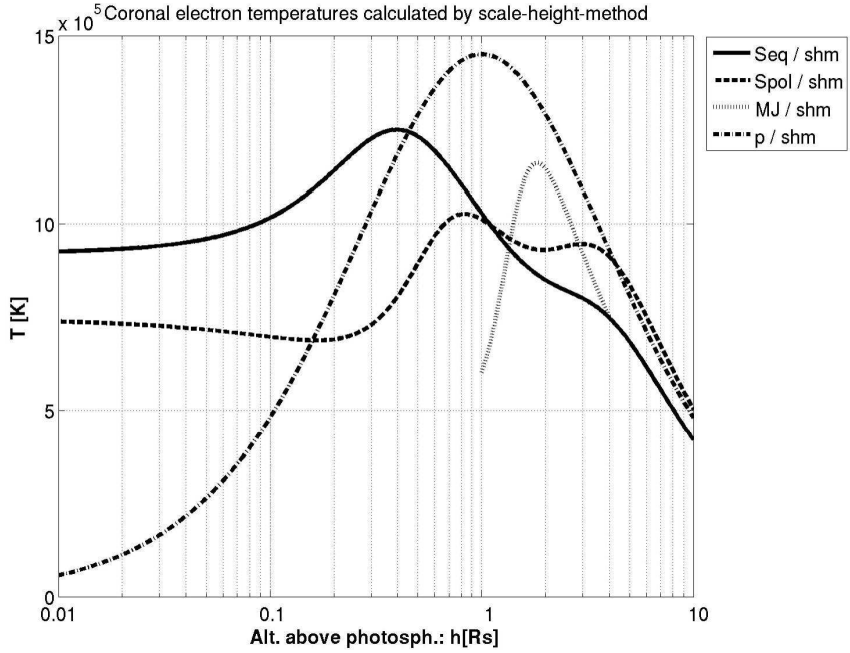

Figure 2. *Coronal electron temperature distributions determined using the scale-height-method (shm)*. The *shm* method is used to calculate electron temperature distributions from the empirical density profiles shown in Figure 1: Saito's equatorial (*Seq*), Saito's polar (*Spol*) density profiles, and *MJ* and *P* (corresponding to Munro and Jackson's and to Pottasch's density profiles, respectively). Note that the electron temperature distributions are not isothermal: they all have a maximum value at r > 1.2 $R_S$, which is well above the transition region, $r_{TR}$ = 1.003 $R_S$. The temperature peak is located at a lower altitude above the equatorial region (*Seq*) than above the polar regions (*Spol*) and in coronal hole models (*MJ*). The values of $T_{max}$ and $h_{max}$ are given in Table 2.

In the equatorial corona (*Seq-shm*), the temperature maximum is at $h_{max}$ = 0.4 $R_S$: $r_{max}$ = 1.4 $R_S$. Above the polar regions (*Spol-shm*) and in Munro and Jackson's coronal hole (*MJ-shm*), the temperature maximum is located at higher altitudes, at $r_{max}$ = 1.83 $R_S$ and 2.8 $R_S$, respectively.

3. **The hydrostatic method (*hst*)**

Since Figure 2 indicates that the corona is not isothermal, the barometric formulae (1) and (4) are not solutions of the equation of hydrostatic equilibrium (3). Both formulae are derived from the tacit assumption that dT/dr = 0 in the solar corona, which is not the case.

When the corona is effectively in hydrostatic equilibrium, T(r) is a solution of the following first-order ordinary differential equation:

$$dT/dr + T\,(d\,Ln\,n/\,dr) = -\mu\,m_H\,g_o\,R_s/k\,r^2 \qquad (7)$$

The *shm* solution (6) is achieved only when and where dT/dr = 0. Otherwise, where dT/dr is negative (i.e., at large radial distances), the solutions of eq (7) necessarily give values for T(r) that are smaller than those provided by the expression (6). Conversely, closer to the Sun where dT/dr > 0, the solutions of eq (7) lead to values for T(r) that are larger than those given by the *shm* expression (6). As a result, actual temperature peak solutions of eq (7) will always be located at a lower altitude than those obtained by the *shm*-method for the same density profile.

The solutions of the differential eq (7) are labelled *hst* (stands for 'hydrostatic equilibrium method'). It is important to note that the *hst* solutions illustrated in the following plots differ significantly from those of eq (6) labelled *shm* and shown in Figure 2. The numerical method used to calculate *hst* solutions is explained below.

*An 'intrinsic potential temperature': $T^*$.* Alfvén [1] introduced this in his 1941 paper, with $T^*$, defined as:

$$T^* = G M_S m_H \mu / 2 k R_S \tag{8}$$

$2 k T^*$ corresponds to the gravitational potential energy of a particle of mass ($\mu m_H$) on the surface of the Sun or a star whose radius is equal to $R_S$ and whose mass is equal to $M_S$. In the case of the Sun, $T^* = 17$ MK, for $\mu = 0.5$.

Note that the *shm* temperature (6) corresponds to $2 T^* / [-d \ln n_e / d(1/r)]$. This means that the coronal temperature calculated by the *shm* method is also proportional to $T^*$.[7]

Alfvén defined a dimensionless temperature variable, $y = T/T^*$, which allows eq (7) to be simplified:

$$dy/dr + y \, (d \ln n / dr) = -1/r^2 \tag{9}$$

The solution of (9), for which $y(r) = 0$ when $r \to \infty$, has the remarkable analytic form:

$$y(r) = - [1 / n(r)] \int_\infty^r [n(r') / r'^2] \, dr' \tag{10}$$

It can be checked that the definite integral is indeed the solution of the first order differential equation (9) for which $y(r \to \infty) = 0$. Because the definite integral in eq (10) tends to zero when $r$ increases to infinity, it is the unique solution of the hydrostatic equation (7) for which $T(r \to \infty) = 0$.[8]

In the next section, the *hst* temperature distributions are calculated by eq (10) for several of the density profiles shown in Figure 1. The differences with the *shm*-solutions obtained by the scale-height-method are emphasized below.

In 1960, Pottasch [26] pointed out that the temperature distribution that satisfies the hydrostatic eq (7) differs basically from that determined by the simple *shm*-method (6), but his warning was not heeded in subsequent decades. This might have been because he did not explain clearly how he solved eq (7) and obtained a maximum temperature of 1.43 MK between r = 1.1 $R_S$ and 2 $R_S$ (see Pottasch's Table 1, and Figure 2). Pottasch's calculation did not use solution (10), nor did he refer to the work of Alfvén [1] [9].

---

[7] Note that the *intrinsic potential temperature* $T^*$ should be seen here as a simple normalizing factor that happens to be of the same order of magnitude as the central temperature of the Sun (13 MK). The ratio of the gravitational potential energy ($G M_S m_H / R_S$) of a proton at the base of the coronal, and the kinetic energy ($k T^*$), is comparable to $\lambda$, the dimensionless parameter introduced by Parker [25] in his hydrodynamic theory of solar wind expansion.

[8] From a historical point of view it is worth recalling that Alfvén [1] published his contribution in a Scandinavian Journal in 1941, a year before Edlén [16] published his well-known identification of coronal emission lines that led him to conclude that the solar corona is an ionized gas with a temperature exceeding 1 MK. The maximum temperature that Alfvén found at $r > 1.2 R_S$ was 1.98 MK. It is surprising that his ingenious method [eq (10)] for calculating coronal temperatures from WL coronal brightness measurements has been overlooked for so long.

It should be noted here that analytical expressions such as (10) offer a key advantage for calculating coronal temperatures, especially when n(r) is fitted by a sum of power functions of 1/r. In this case, the definite integral in eq (10) can be determined analytically, as in the original work by Alfvén [1].

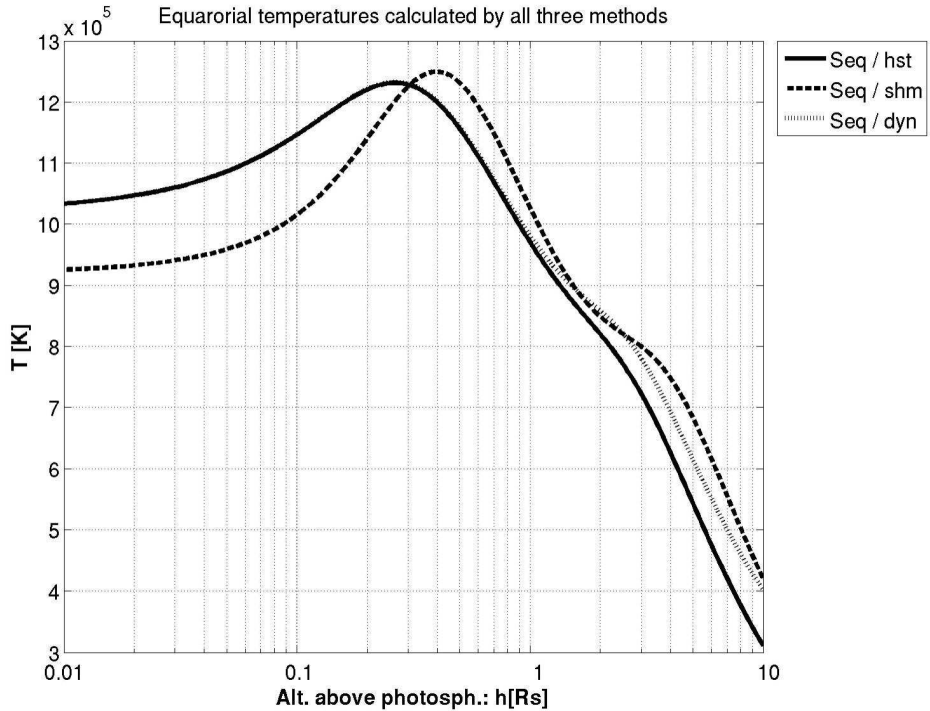

Figure 3. *Coronal electron temperature distributions derived using different methods of calculation (shm, hst and dyn).* The same empirical electron density distribution (*Seq*) is used for all cases; *shm* corresponds to the scale-height-method as in Figure 2; *hst* corresponds to the hydrostatic equilibrium method introduced by Alfvén [1] in 1941; and *dyn* corresponds to the dynamic or hydrodynamic equilibrium method developed to calculate electron temperature distributions when the solar corona is expanding hydrodynamically. The temperature peaks calculated using the *hst* and *dyn* methods are comparable to those calculated using the *shm* method, but they are formed at lower altitudes. The *hst* and *dyn* methods give almost the identical temperatures at the lower altitudes where the hydrodynamic expansion velocity is small (subsonic). At higher altitudes, where the plasma is accelerated to supersonic velocities, larger differences are obtained.

*Coronal temperature distributions determined by the hst method.* We focus first here on the equatorial density profile derived from Saito's [27] empirical model illustrated in Figure 1 by the thick solid line labelled *Seq*. The coronal temperature distribution obtained using the *shm-*method is indicated by the *Seq/shm* curve in Figures 2 and 3. The temperature profile obtained using Alfvén's method or the *hst* method is illustrated by the *Seq/hst* curve for the same equatorial density profile [10].

---

[9] Pottasch [26] was probably unaware of the remarkable analytic solution (10) discovered by Alfvén two decades earlier. Indeed, the way he calculated the solution of the hydrostatic equation is not clear because he does not specify an initial condition to solve the first-order differential eq (7). He claimed to have used an "iterative numerical method" that "is reasonably rapidly converging", but without additional explanations. Note that the shape of the coronal temperature profile he published differs considerably from the shape we obtained from eq (10) for the same density profile (*P*) shown in Figure 1.

From Figure 3 it can be seen that the maximum temperature obtained using the *hst* method is almost equal to that obtained when the *shm* method was used. Indeed, where temperature is maximum $dT/dr = 0$ eq. (9) is equivalent to eq. (3) and the solution (10) equivalent to the solution (6). More importantly, however, it can be seen that $h_{max}$, the altitude corresponding to the maximum temperature, is significantly smaller ($h_{max} \approx 0.26\ R_S$) using the *hst* solution than the *shm* one ($h_{max} \approx 0.4\ R_S$). The reason for this remarkable difference lies in presence of the first term in eq (7). Similar conclusions hold for the other electron density profiles shown in Figure 1.

Since the altitude of the maximum coronal temperature is related to the altitude where the heating rate of coronal plasma is greatest, it is essential to use a reliable method to calculate coronal electron temperature distribution: i.e. eq. (10), or the more general formula (15) derived in the next section.

Indeed, a third method is presented there for the more general case when the corona is not in hydrostatic equilibrium, but is expanding hydrodynamically, as discovered by Parker [25]. Indeed the continuous radial expansion of the solar corona is now a well-established observed phenomenon.[11]

## 4. The hydrodynamic method (*dyn*)

Since the solar corona is expanding, hydrostatic eq (3) needs to be replaced by the momentum transport equation; the inertial force term should then be added to the gravitational force, and to pressure gradient in the left hand side of eq (3). This led us to search for a third method to determine the radial distribution of the coronal temperature which is a solution of the hydrodynamic equations.
This third method is labelled *dyn (*stands for 'dynamic' or 'hydrodynamic', distinguishing it from Alfvén's 'hydrostatic equilibrium method', labeled *hst*).

Taking u(r) to be the radial component of the solar wind bulk velocity, let us consider that the coronal expansion is stationary and that the flux of mass is conserved along magnetic flux tubes or flow tubes whose cross-section is a prescribed function of r. The function introduced by Kopp and Holzer [21] has been adopted by many solar wind modellers to approximate the geometrical expansion rate of coronal flow tubes:

$$A(r) / A(R_S) = r^2\ f(r) = r^2\ \{f_{max}\ \exp[(r - r_1)/\sigma] + f_1\} / \{\exp[(r - r_1)/\sigma] + 1\} \qquad (11)$$

---

[10] The electron kinetic pressure, $p_e = n_e\ k\ T_e$, satisfies the hydrostatic eq (7) when the *hst* solution for $T_e(r)$ is used in association with Saito's equatorial densities for $n_e(r)$. When *shm* temperature distribution is used with Saito's equatorial density distribution, however, it can be shown that $p_e(r)$ fails to satisfy the hydrostatic eq (7).
[11] Parker's well-known argument in support of the radial expansion of solar wind was based on the imbalance of the mechanical pressures between interplanetary space and the base of the solar corona. On the other hand, Lemaire [34] showed that any hydrostatic equilibrium model of the solar corona (including Chapman's conductive-hydrostatic model [10]) becomes convectively unstable above a certain altitude. This implies that the coronal plasma is expanding hydrodynamically into interstellar space because the heat conduction is simply not efficient enough to evacuate the excess energy deposited in the inner corona near where the coronal temperature is at its maximum.

where $f_{max}$ is the maximum super-radial expansion rate; $r_1$ is the heliocentric distance of the steepest expansion rate; $\sigma$ is the normalized range over which this faster expansion rate is confined; and $f_1 = 1 + (1 - f_{max}) \exp[(1 - r_1)/ \sigma ]$; $f(1) =1$ on the surface of Sun. [12]

When $f_{max} = 1$, the usual radial expansion rate of flow tubes is recovered, as assumed in early solar wind models. In applying the 'hydrodynamic method' of calculation, Brandt et al. [8] assumed such a radial solar wind expansion. In Munro and Jackson's [24] study of a polar coronal hole, the following values were adopted for these input parameters: $f_{max} = 7.26$, $r_1 = 1.31$, and $\sigma = 0.51$. Cranmer et al. [11] parameterized the coronal hole expansion rate by $f_{max} = 6.5$, $r_1 = 1.5$, and $\sigma = 0.6$, whereas the expansion factor adopted by Deforest et al. [12] can be fitted by $f_{max} = 5.65$, $r_1 = 1.53$, and $\sigma = 0.65$.

All these fits are quite similar. In our study, we used $f_{max} = 1$ to simulate radial expansion, as well as a series of other values listed in Table 1. Smaller and larger values of $f_{max}$, $r_1$ and $\sigma$ were also used in the following investigation of the influence of these geometrical parameters on the distributions of coronal temperatures calculated using the *dyn* method.

Let us assume that the coronal plasma is quasi-neutral and that it is homogeneous with $\mu = 0.5$. The effect of $He^{++}$ ions concentration will be addressed later. Let us also assume that the outward bulk velocity of the electrons, $u_e$, is equal to the bulk velocities of all ions species, $u_i$ (i.e., that there is no net radial electric current, nor diffusion of any particle species with respect to the others [diffusive equilibrium]). Based on these hypotheses (restrictions/limitations), the coronal plasma can be treated as a single MHD fluid whose bulk velocity, $u(r) = u_e = u_i$, is a solution of the mass and momentum flux conservation equations.

The electron and ion temperatures vary with r, but it is assumed that their ratio, $T_p/T_e = \tau_p$, is independent of r. Although this assumption might have to be revisited in future simulations, it is made here in order to understand the first-order effect of changing this input parameter. In the following set of calculations, it is assumed that $\tau_p = 1$. Other values are assumed later. In addition, the total kinetic pressure of the solar corona is assumed to be isotropic (i.e., the velocity distribution functions of the protons and electrons are isotropic within the frame of reference co-moving with their bulk speed $u(r)$ [13].

Obviously, all these simplifying assumptions need to be viewed as attempts to investigate, step by step, the role of each hypothesis about the calculated distribution of coronal temperatures. From the conservation of mass flow, the steady state distribution of u(r) can be easily obtained from the given density profile:

$$u(r) = u_E [A_E / A(r)] [n_E / n(r)] \quad (12)$$

where $A_E$, $u_E$, and $n_E$ are the cross-section of the flow tube, the solar wind bulk velocity, and the electron density at 1AU, respectively. Here, the values of $u_E$ and $n_E$ are free input parameters sometimes taken from averaged measurements at the orbit of Earth. For the quiet solar wind, $n_E = 5.65$ electrons/cm$^3$, and $u_E = 329$ km/s. For fast solar wind streams, $n_E = 2.12$ electrons/cm$^3$ and $u_E = 745$ km/s [17]. Other values adopted for these input parameters are listed in Table 1 for each model calculation.

---

[12] An alternative expansion rate of flow tubes was reported by Wang et al. [ 40]. It is qualitatively similar, however, to that introduced by Kopp and Holzer [21] and adopted here.
[13] The limitations of this assumption are discussed in Appendix 2 of a review paper by Echim et al. [15].

The differential equation that has to be solved to determine the coronal electron temperature that satisfies the hydrodynamic momentum equation is derived in Appendix A [14]. As in the previous section, the dimensionless temperature variable, $y = T_e / T^*$, is used. The hydrodynamic momentum equation is then:

$$dy/dr + y \; d \, Ln \, n/dr = - 1/r^2 \; [1 + F(r)] \qquad (13)$$

where $F(r)$ is the ratio of the 'inertial acceleration' of the solar wind over its 'gravitational deceleration', $g(r)$:

$$F(r) = r^2 \, u \, (du/dr) / g_o \, R_S \qquad (14)$$

Note that this function of r is fully determined by the values of $u(r)$ and $du(r)/dr$ that can be derived from $n(r)$ using eqs (12) and (5) or (16).

The $y(r)$ solutions are obtained by integrating numerically the first-order differential eq (13). A boundary condition, however, must be met for $y$ at some reference level: either at $r = r_b$, or at $r = \infty$. Brandt et al. [8] chose initial conditions at $r_b = 1$, the base of the corona. Unfortunately, unless the initial condition $y_b$ is fixed with a large number of digits, the numerical solutions of eq (13) generally diverge when $r \rightarrow \infty$. This forced these authors to adjust the value of $y_b$ very precisely, through trial and error, until a "well-behaved solution" was eventually identified for $y(r)$[15]. The family of diverging solutions generated by this iterative procedure is illustrated by Figure 2 in Brandt et al.'s paper. More than 5 or 6 digits have to be given for this "well-behaved solution" to extend beyond $r = 10$. It is probably because of this mathematical inconveniency that Brandt's iterative method for calculating the coronal temperature has not been adopted by other solar physicists, apart from Munro and Jackson [24] and perhaps a few others.

An alternative and more straightforward procedure is proposed here. Without any cumbersome iterative procedure, it gives the expected solution of (13) tending asymptotically to zero when $r \rightarrow \infty$:

$$y(r) = \; T(r) \; / \, T^* = - [1 / n(r)] \int_\infty^r [n(r') / r'^2] \, [1 + F(r')] \, dr' \qquad (15)$$

$T^*$ is the *intrinsic potential temperature* already defined by eq (8) or more generally by (A11). As in Alfvén's work, $T^*$ is a convenient normalizing factor for the coronal temperature. The definite integral in eq (15) can be calculated by using a standard integration algorithm because $n(r)$ and $F(r)$ are known functions of r.

---

[14] Note that in usual hydrodynamic solar wind model calculation, the momentum transport equation is integrated to determine the distribution of the bulk velocity $u(r)$, and not the temperature distribution $T(r)$. But because the equation of conservation of the mass flow (12) uniquely determines $u(r)$ as a function of r, the momentum equation can be used to obtain $T(r)$, as done by Alfvén [1], who integrated the hydrostatic equation to obtain $T(r)$. Note that when the distributions of n, u and T are determined by the *dyn* method as explained here, the energy transport equation can be employed to determine the coronal energy deposition rate (i.e., the heating rate within the corona). The latter is often guessed (arbitrarily postulated) in standard hydrodynamic solar modeling applications [15].

[15] Note that a similar divergence is obtained near the 'critical point' of the family of solar wind velocity $u(r)$, which are solutions of the hydrodynamic momentum equation, in standard solar wind models.

*TABLE 1. Input parameters used in eqs (11) – (16) to define the density profiles illustrated in Figure 1, which are used to calculate the temperature distributions (6), (10) and (15). The second column corresponds to the model ID, also used in Table 2, and in Figures 2, 3, 4 and 5. NB: Some of the models listed here are given for reference, but are not shown in the figures for brevity.*

| # | Model ID | Latitude | $n_E$ [cm$^{-3}$] | $u_E$ [km/s] | $f_{max}$ - | $r_1$ [$R_S$] | $\sigma$ [$R_S$] | $T_p/T_e$ - | $n_{He}/n_p$ - | References and key parameters |
|---|---|---|---|---|---|---|---|---|---|---|
| 1 | Seq | **0°** | **5.75** | **329** | 1 | - | - | 1 | 0.00 | Saito [27] equator with SW ext. / $u_E$=329km/s |
| 2 | Spol | **90°** | **2.22** | **600** | 1 | - | - | 1 | 0.00 | Saito [27] pole with SW ext./ $u_E$=600km/s |
| 3 | MJ | **90°** | **2.22** | **600** | 1 | - | - | 1 | 0.00 | Munro and Jackson [24] CH / $u_E$=600km/s |
| 5 | P | 0° | 5.10 | 300 | 1 | - | - | 1 | 0.00 | Pottasch [26] equator / $u_E$=300km/s |
| 6 | Spz | 90° | **2.22** | **1** | 1 | - | - | 1 | 0.00 | Saito [27] pole without SW / $u_E$=1km/s |
| 7 | Spv100 | 90° | **2.22** | **100** | 1 | - | - | 1 | 0.00 | Saito [27] pole./ $u_E$=100km/s |
| 8 | Spv | 90° | 2.22 | **329** | 1 | - | - | 1 | 0.00 | Idem / $u_E$=329km/s |
| 9 | Spv450 | 90° | 2.22 | **450** | 1 | - | - | 1 | 0.00 | Idem / $u_E$=450km/s ; $f_{max}$=1 |
| 10 | Spvd | 90° | 2.22 | 329 | **0.8** | 1.31 | 0.51 | 1 | 0.00 | Idem ./ $u_E$=329km/s; **$f_{max}$=0.8** |
| 11 | SpvD | 90° | 2.22 | 329 | **3** | 1.31 | 0.51 | 1 | 0.00 | Idem / $u_E$=329km/s; **$f_{max}$=3** |
| 12 | SpvD5 | 90° | 2.22 | 329 | **5** | 1.31 | 0.51 | 1 | 0.00 | Idem ./ $u_E$=329km/s; **$f_{max}$=5** ; $r_1$=1.31 |
| 13 | SpvDh | 90° | 2.22 | 329 | 3 | **2.00** | 0.51 | 1 | 0.00 | Idem / $u_E$=329km/s; $f_{max}$=3; **$r_1$=2.00** |
| 14 | SpvDh3 | 90° | 2.22 | 329 | 3 | **3.00** | 0.51 | 1 | 0.00 | Idem / $f_{max}$=3; **$r_1$=3.00**; $\sigma$=0.51 |
| 15 | SpvDhw1 | 90° | 2.22 | 329 | 3 | **2.00** | **1.00** | 1 | 0.00 | Idem / $f_{max}$=3; $r_1$=2.00; **$\sigma$=1** |
| 16 | SpvDh2w2 | 90° | 2.22 | 329 | 3 | 2.00 | **2.00** | 1 | 0.00 | Idem / $f_{max}$=3; $r_1$=2; **$\sigma$=2** |
| 17 | SpvDw1 | 90° | 2.22 | 329 | 3 | **1.31** | **1.00** | 1 | 0.00 | Idem / $f_{max}$=3; $r_1$=1.31; **$\sigma$=1** |
| 18 | SpvDw2 | 90° | 2.22 | 329 | 3 | 1.31 | **2.00** | 1 | 0.00 | Idem / $f_{max}$=3; $r_1$=1.31; **$\sigma$=2** |
| 19 | SpvD5 | 90° | 2.22 | 329 | **5** | 1.31 | 0.51 | 1 | **0.00** | Idem / **$f_{max}$=5**; $r_1$=1.31; **$\sigma$=0.51** |
| 20 | SpvHe5p | 90° | 2.22 | 329 | 1 | - | - | 2 | **0.05** | Idem / $f_{max}$=1; **$n_{He}/n_p$= 5%**; $T_p/T_e$=2 |
| 21 | SpvHe10p | 90° | 2.22 | 329 | 1 | - | - | 1 | **0.10** | Idem / $f_{max}$=1; **$n_{He}/n_p$= 10%**; $T_p/T_e$=1 |
| 22 | SpvHe20p | 90° | 2.22 | 329 | 1 | - | - | 2 | **0.20** | Idem / $f_{max}$=1; **$n_{He}/n_p$= 20%**; $T_p/T_e$=2 |
| 23 | SpvTp2 | 90° | 2.22 | 329 | 1 | - | - | **2** | 0.00 | Idem / $f_{max}$=1; $n_{He}/n_p$= 0%; **$T_p/T_e$=2** |
| 24 | SpvTp4 | 90° | 2.22 | 329 | 1 | - | - | **4** | 0.00 | Idem / $f_{max}$=1; $n_{He}/n_p$= 0%; **$T_p/T_e$=4** |

The dotted curve labelled (*Seq/dyn*) in Figure 3 shows the equatorial temperature distribution obtained from (15) for Saito's extended density model that is defined by:

$$n(r) = 10^8 [3.09 \, r^{-16} (1 - 0.5 \sin \Phi) + 1.58 \, r^{-6} (1 - 0.95 \sin \Phi) + 0.0251 \, r^{-2.5} (1 - \sin^{0.5} \Phi)] + n_E (215/r)^2 \quad (16)$$

where $\Phi$ is the heliographic latitude. The first terms of this sum correspond to Saito's [27] averaged 2-D coronal density distribution. The last term has been added to extend the coronal density distribution beyond 10 $R_S$; it corresponds to the asymptotic solar wind density at large distances where u(r) is supersonic and almost independent of r.

The constant $n_E$ is the electron density measured in the solar wind at 1AU. For the equatorial corona ($\Phi = 0$), we used the following values: $n_E = 5.65$ electrons/cm$^3$, $u_E = 329$ km/s, taken from Ebert et al. [17] for the quiet/slow solar wind at 1AU in the ecliptic plane.

The dotted curve (*Seq/dyn*) in Figure 3 gives the temperature distribution calculated using the *dyn* method for this equatorial density model. It can be compared with the distribution obtained using the other two methods (*shm* and *hst*) for the same equatorial density distribution (16) [16].

A comparison of the solid and dotted curves, *Seq/hst* and *Seq/dyn*, respectively, indicates that a slow solar wind expansion velocity does not drastically change the temperature distribution in the inner corona where the radial expansion velocity is small (subsonic). Larger differences appear, however, beyond $h_{max}$, the altitude where the temperature reaches its maximum value. It can be seen that, in the equatorial region of the corona, $h_{max}$, is almost identical for the *dyn* and *hst* solutions. Note that $h_{max}$ differs significantly from that obtained with the more commonly used *shm* method.

Figure 4 shows the temperature distributions obtained using the *dyn* method for other density models. The curves in Figure 4 differ significantly from those shown in Figure 2 for the *shm* method. For each temperature profile, the values of $T_{max}$, and $h_{max}$ are given in Table 2 for both the *dyn* and *shm* methods. The width of the temperature peaks is also given in Table 2 for the *dyn* temperature distributions. The latter will be defined and discussed in a following subsection; it should be seen as a measure of the range of altitudes over which the coronal heating source extends.

From Table 2 and Figure 4 it can be concluded that the temperature distributions in the polar regions and in coronal holes differ significantly from those corresponding to the equatorial region. In addition, over the polar regions the temperature profile appears to be much more sensitive to coronal expansion velocity than over the equatorial region, where coronal density is relatively larger.

These results lead to the interesting conclusion that the mechanisms of coronal heating are distributed quite differently over the polar and equatorial regions. In addition, the energy deposition rate appears to be spread over a much wider range of radial distances ($\Delta r = 2 – 4 \, R_S$) in the polar regions than in the equatorial region, where $\Delta r = 0.4 – 0.6$ $R_S$. Also, the peak of temperature is closer to the base of the corona at low latitudes.

---

[16] When $u_E = 0$, the hydrostatic equilibrium is recovered and the *dyn* method then gives the same results as the *hst* method; according to eqs (12) and (14), in this case u(r) = 0, F(r) = 0 and the solutions (10) and (15) are then identical.

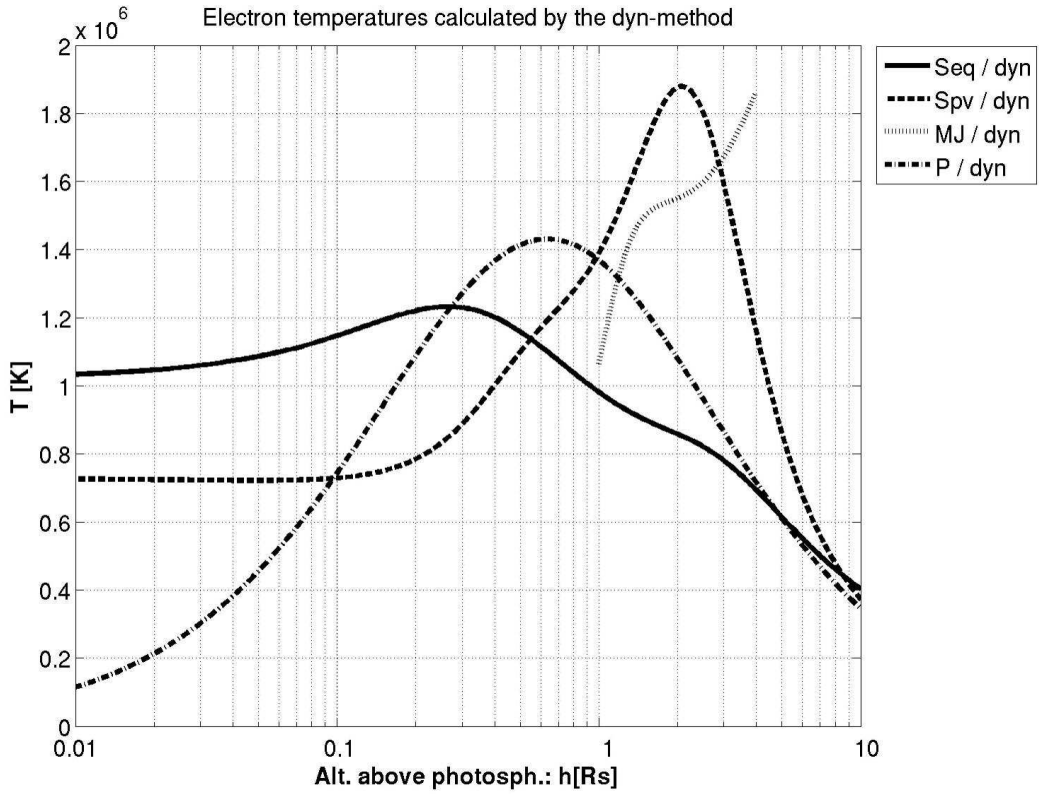

Fig 4. *Coronal electron temperature distributions calculated using the dyn method [eq 15]*. The input parameters for the electron density models (*Seq; Spv; MJ; P*) are listed in Table 1. The calculated temperature maximum (T'$_{max}$), altitude (h'$_{max}$) and half-width range ($\Delta r$) are given in Table 2 for each case. The values of T'$_{max}$ obtained using the *dyn* method are not equal to those obtained using the *hst* method unless solar wind expansion velocity at 1 AU is smaller than 150 km/s. In addition, the altitude of the temperature peaks, h'$_{max}$, determined using the *dyn* method, are higher up in the corona than those obtained using the *hst* one (see Table 2). Therefore, to determine the altitudes where the maximum energy is deposited to heat the solar corona, the *dyn* method should be used instead of the *hst* method.

*Effect of coronal expansion velocity.* Figure 5 shows a set of temperature distributions obtained for the Saito's extended polar region average density model, taking account of solar wind expansion velocities gradually changing from $u_E = 1$ to 450 km/s at 1AU. The five curves correspond to $u_E = 1$ km/s, 100 km/s, 329 km/s and 450 km/s for *Spz*, *Spv100*, *Spv*, and *Spv450*, respectively. All these temperature distributions are calculated for a radial expansion rate ($f_{max} = 1$) and for $n_E = 2.12$ cm$^{-3}$, the average electron density of the fast solar wind at 1AU. It can be seen that the maximum temperature is a very sensitive function of $u_E$.

For $u_E > 150$ km/s, the peak of temperature exceeds 1 MK and is located at $r_{max} \approx 3$. This altitude is where $u.du(r)/dr$, the acceleration rate of the coronal expansion velocity, is steepest; this is where the function F(r) reaches a maximum value. The maximum, $F_{max}$, increases with $u_E$. The value of the definite integral in (15) therefore increases with $u_E$. The rapid growth of T(r) is therefore a result of the steep increase of the function F(r) at $r \approx r_{max}$, where $u(r).du(r)/dr$ reaches its maximum value. This effect becomes prominent when $u_E > 150$ km/s.

For $u_E$ < 150 km/s, Figure 5 shows that there are two separate temperature peaks. The first one, $T_{max}$, is located nearest the Sun at $r_{max} \approx 1.5 - 1.7$ (i.e., almost where the temperature maximum was found with the *hst* method). [17]

The second temperature peak, $T'_{max}$, is located at far higher altitudes : r > 3. As indicated earlier when the value of $u_E$ becomes greater than about 150 km/s, this second temperature peak becomes the more prominent one, and develops at $r'_{max} \approx 3.2$ where F(r) is maximum in the case of Saito's extended polar density model.

When $u_E$ = 329 km/s, the second temperature maximum is equal to 1.8 MK (see *Spv/dyn* curve). When $u_E$ = 450 km/s, the *Spv450/dyn* curve reaches a maximum of $T'_{max}$ = 2.75 MK. Note that these values are much greater than $T_{max}$ obtained using the *hst* and *shm* methods.

When a solar wind velocity of $u_E$ = 600 km/s is applied, the *dyn* method gives a peak temperature of 4.1 MK. This maximum is reached at r = 3.1 over the polar regions for Saito's extended polar density model, when $f_{max}$ = 1 is assumed in the calculation (this curve is not shown in Figure 5). The straightforward correlation between $T'_{max}$ and $u_E$ associated with a particular coronal density profile can be used as a future diagnostic tool to determine the dependence (and the correlation coefficient) of solar wind velocities on the energy deposition rates in the inner corona.

Note that coronal hole electron temperatures were derived from measurements of intensity ratios of collisionally excited EUV and visible spectral lines [37] [38]; they are less than 1.2 MK and 2.2 MK, respectively. They correspond, however, to observations made at r < 1.5 – 1.6 (i.e., at much lower altitudes than where the temperature peak is found using the *dyn* method over the poles (for large expansion velocities).

The range of altitudes where the coronal electron temperature reaches a maximum is where the coronal plasma is steeply accelerated and boosted to supersonic velocity. It is also where the coronal heating rate is at its maximum. This correspondence is qualitatively consistent with the concept that faster solar wind velocities at 1AU require higher maximum temperatures in the inner corona, and therefore larger coronal heating rates.

Figure 5, as well as the evidence that the calculated temperature distribution over the polar regions is very sensitive to the coronal expansion velocity, infers that the *dyn* method should be generally adopted to determine coronal temperatures from WL brightness measurements. This is because the *hst* and *shm* methods produce misleading results at r = 3 and beyond; also, they ignore the coronal expansion rate, and the second (the highest) temperature peak, $T'_{max}$, associated with high supersonic expansion speeds can be obtained only when the *dyn* method is used.

*Width of the temperature peak.* The range of altitudes over which T(r) is greater than $T_{max}/2$ or $T'_{max}/2$ is reported in Table 2 in column Δr. The values of Δr are given in $R_S$ units for each temperature distribution calculated using the *dyn* method. As mentioned earlier, this range of altitudes can be related to the extent to which maximum energy is deposited to heat the coronal plasma and to accelerate it to maximum supersonic velocities.

---

[17] The solid line labelled *Spz* has two temperature maximae; it corresponds to Saito's polar density model with a solar wind density of $n_E$ = 2.22 e/cm$^3$, and a negligibly small expansion velocity: $u_E$ = 1 km/s. The contributions of solar wind velocity, u(r), and acceleration, du/dr, to F(r) and to y(r), are then negligibly small; the *Spz/dyn* temperature distribution is therefore almost identical to the *Spz/hst* one, as in the case of hydrostatic equilibrium.

The empirical values of Δr and h'$_{max}$ derived from WL eclipse observations should be seen as constrains that existing and future theories of coronal heating will need to match. Electron temperature distributions such as those illustrated in Figures 4 and 5 are therefore benchmarks for specifying which energy sources and dissipation mechanisms best account for heating the inner solar corona, for accelerating the solar wind to supersonic speeds, and for the electron density profiles observed during total solar eclipses.

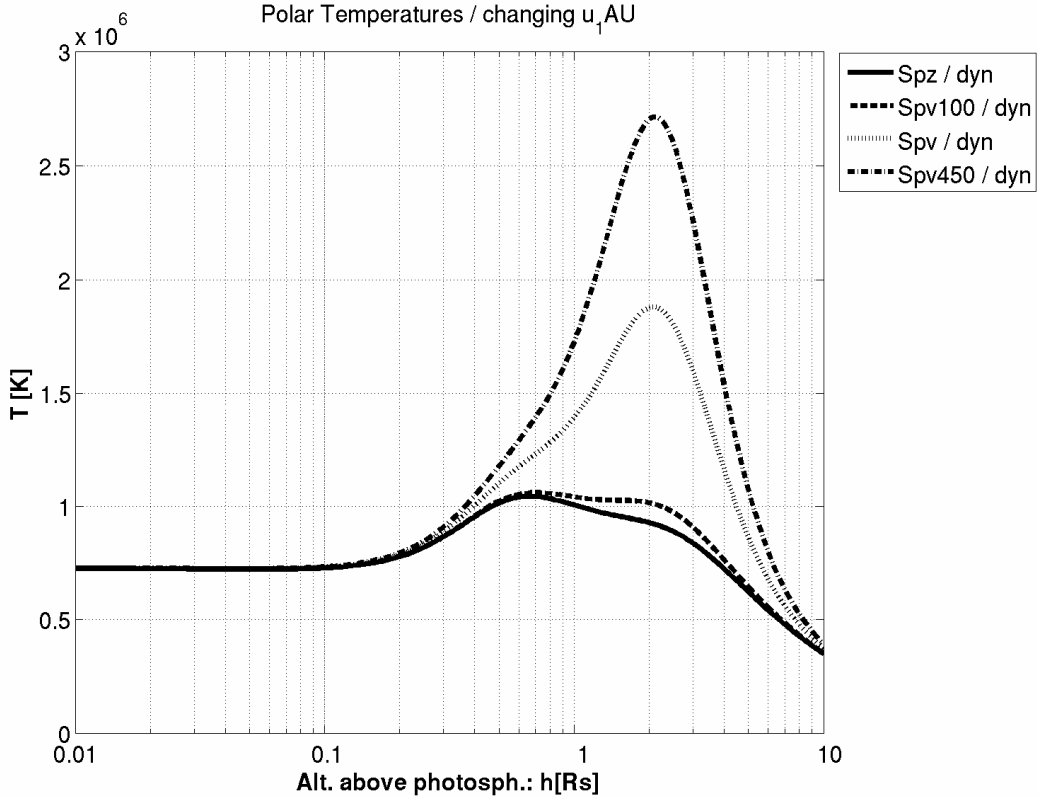

Figure 5. *Effect of coronal expansion velocity*. The *dyn* temperature distributions derived from Saito's [27] average polar coronal density model extended to match solar wind density at 1AU ( $n_E$ = 2.22 cm$^{-3}$), and different values for the bulk velocity ( $u_E$ = 1 / 100 / 329 / 450 km/s), are illustrated by the *Spz, Spv100, Spv*, and *Spv450* curves, respectively. Radial flow is assumed for all these cases (f$_{max}$ = 1). For $u_E$ > 150 km/s, a new temperature peak T'$_{max}$ forms at higher altitude, r ≈ 3 R$_S$. This is where the coronal expansion velocity has its steepest acceleration rate; that is, where F(r) has a maximum (see eq 14). The value of T'$_{max}$ increases with $u_E$. These numerical results confirm that larger solar wind velocities at 1AU require higher temperatures and larger energy deposition rates in the inner corona.

Coronal temperature distributions derived from empirical electron density distributions using the *dyn* method could therefore contribute to resolving these questions: What are the predominant physical mechanisms that heat the solar corona? What are the heating sources that account for the electron density distributions derived from WL brightness observations, such as eq (16)? To answer these questions, it is still necessary to find a way to link the values of $n_E$ and $u_E$ at 1AU, with coronal density profiles derived in the inner corona from WL observations. Coordinated spacecraft measurements of $n_E$ and $u_E$ at 1AU, and coronal electron temperatures inferred from other types of observations, might help to meet this challenge. The coronal eclipse observations of Fe X and Fe XIV emission lines with a narrow interference filter reported by Habbal et al. [37] could be useful in this respect. But this question is beyond the scope of our current study.

*TABLE 2. The largest of the electron temperature peaks ($T_{max}$, $T'_{max}$) calculated by the dyn method is given in 3$^{rd}$ column; the altitude of the corresponding maximum temperature ($h_{max}$, $h'_{max}$) is given in 4$^{th}$ column; the width of the corresponding temperature peak($\Delta r$) is given in 5$^{th}$ column for some of the density profiles listed in Table 1. The temperature maximum and altitude calculated by the scale height method (shm) are given in columns 6 and 7 for comparison. The last column contains refs, and characteristic input parameters.*

| # | Model ID | $T_{max}$ or $T'_{max}$ MK (dyn) | $h_{max}$ or $h'_{max}$ $R_S$ (dyn) | $\Delta r$ $R_S$ (dyn) | $T_{max}$ MK (shm) | $h_{max}$ $R_S$ (shm) | References and key parameters |
|---|---|---|---|---|---|---|---|
| 1 | Seq | 1.23 | 0.26 | 0.47 | 1.25 | 0.40 | Saito [27] equator with SW ext. |
| 2 | Spol | 4.11 | 2.14 | 2.70 | 1.03 | 0.83 | Saito [27] ) pole with SW / $u_E$=600km/s |
| 5 | P | 1.43 | 0.63 | 3.47 | 1.45 | 1.00 | Pottasch [26] equator with SW / $u_E$=300km/s |
| 6 | Spz | 0.94 | 0.56 | 0.67 | 0.95 | 0.74 | Saito [27] pole without SW / $u_E$=1km/s |
| 7 | Spv100 | 1.06 | 0.71 | 2.76 | 1.03 | 0.83 | Saito [27] pole with SW / $u_E$=100km/s |
| 8 | Spv | 1.88 | 2.09 | 2.78 | 1.03 | 0.83 | Idem / $u_E$=329km/s |
| 9 | Spv450 | 2.71 | 2.09 | 2.71 | 1.03 | 0.83 | Idem / $u_E$=450km/s ; $f_{max}$=1 |
| 10 | Spvd | 1.89 | 2.09 | 2.75 | 1.03 | 0.83 | Idem / $u_E$=329km/s; **$f_{max}$=0.8** |
| 11 | SpvD | 1.84 | 2.09 | 2.87 | 1.03 | 0.83 | Idem / $u_E$=329km/s; **$f_{max}$=3** |
| 12 | SpvD5 | 1.84 | 2.09 | 2.89 | 1.03 | 0.83 | Idem / $u_E$=329km/s; **$f_{max}$=5** ; $r_1$=1.31 |
| 13 | SpvDh | 1.78 | 2.09 | 3.07 | 1.03 | 0.83 | Idem / $u_E$=329km/s; $f_{max}$=3; **$r_1$=2.00** |
| 14 | SpvDh3 | 1.70 | 1.07 | 3.08 | 1.03 | 0.83 | Idem / $f_{max}$=3; **$r_1$=3.00**; σ=0.51 |
| 15 | SpvDhw1 | 1.76 | 1.66 | 2.59 | 1.03 | 0.83 | Idem / $f_{max}$=3; $r_1$=2.00; **σ=1** |
| 16 | SpvDh2w2 | 2.14 | 1.51 | 1.78 | 1.03 | 0.83 | Idem / $f_{max}$=3; $r_1$=2; **σ=2** |
| 17 | SpvDw1 | 1.78 | 1.82 | 2.78 | 1.03 | 0.83 | Idem / $f_{max}$=3; $r_1$=1.31; **σ=1** |
| 19 | SpvD5 | 1.84 | 2.09 | 2.89 | 1.03 | 0.83 | Idem / **$f_{max}$=5;** $r_1$=1.31; **σ=0.51** |
| 20 | SpvHe5p | 1.55 | 2.09 | 2.78 | 1.14 | 0.83 | Idem / $f_{max}$=1; **$n_{He}/n_p$= 5%**; **$T_p/T_e$=2** |
| 21 | SpvHe10p | 2.61 | 2.09 | 2.78 | 1.25 | 0.83 | Idem / $f_{max}$=1; **$n_{He}/n_p$= 10%**; **$T_p/T_e$=1** |
| 22 | SpvHe20p | 2.30 | 2.09 | 2.78 | 1.42 | 0.83 | Idem / $f_{max}$=1; **$n_{He}/n_p$= 20%**; **$T_p/T_e$=2** |
| 23 | SpvTp2 | 1.25 | 2.09 | 2.78 | 1.03 | 0.83 | Idem / $f_{max}$=1; $n_{He}/n_p$= 0%; **$T_p/T_e$=2** |
| 24 | SpvTp4 | 0.75 | 2.09 | 2.78 | 1.03 | 0.83 | Idem / $f_{max}$=1; $n_{He}/n_p$= 0%; **$T_p/T_e$=4** |

*The effect of the super-radial expansion of flow tubes.* It has been assumed, so far, that hydrodynamic expansion is strictly radial (i.e., $f_{max}$ = 1), as generally postulated in hydrodynamic or kinetic solar wind models. Let us now investigate the effect of the super-radial expansion of flow tubes generally assumed to coincide with interplanetary magnetic flux tubes connected to the polar regions.

---

[18] Note that this normal assumption is not necessarily applicable when (i) a convection electric field is present in an external magnetic field distribution, and/or (ii) grad-B and curvature drifts are taken into account to determine the motion of charged particles whose kinetic energy is not negligible. Nevertheless, this assumption, commonly made by the MHD community, has also been adopted here, for convenience.

According to eq (16), the distribution of electron densities at large distances is determined by $n_E$, the density at 1AU. The distributions of u(r) and F(r) along flow tubes are determined by the values adopted for the additional free parameters $u_E$, $f_{max}$, $\sigma$ and $r_1$. When $f_{max} > 1$, the cross-section of the flow tube expands faster than $r^2$. This is expected over the poles and coronal holes. Conversely, $f_{max} < 1$ is expected in inter-plume regions, where the flux tubes cross-section varies less rapidly than $r^2$ between $r_1 - \sigma$ and $r_1 + \sigma$.

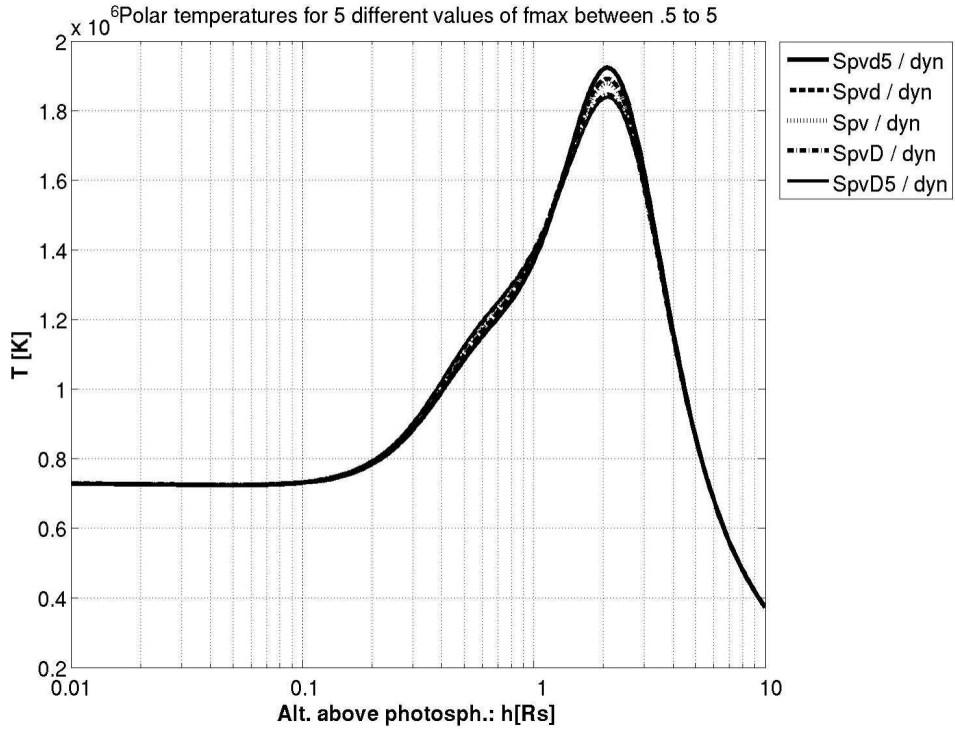

Figure 6. *Effect of super-radial expansion rate.* All these *dyn* temperature distributions are derived from the same density profile: Saito's [27] average polar density model extended to match quiet solar wind density and bulk speed at 1AU ( $n_E$ = 2.22 cm$^{-3}$; $u_E$ = 329 km/s). Non-radial flux tube expansion factors are assumed for all five curves: $f_{max}$ = 0.5 / 0.8 / 1 / 3 / 5 for the *Spvd5, Spvd, Spv, SpvD,* and *SpvD5* curves, respectively.
For all cases: $r_1$ = 1.31, $\sigma$ = 0.51 (see Table 1). It can be seen that the electron temperature maximum (T'$_{max}$) is not a function sensitive to the flux tube expansion factor $f_{max}$, unlike the situation with Kopp and Holzer's [21] polytropic solar wind hydrodynamic model [21].

The effect of changing the value of $f_{max}$, the maximum geometrical expansion, is illustrated in Figure 6 for Saito's polar extended density model *Spol*; the curves labelled *Spvd5*, *Spvd*, *Spv*, *SpvD* and *SpvD5* correspond to $f_{max}$ = 0.5, 0.8, 1.0, 3.0 and 5.0, respectively, all other input parameters being unchanged. The values of all input parameters are listed in Table 1.

It can be seen that the shape of the temperature distributions is not significantly affected by changing, by a factor of 10, the value of $f_{max}$, provided that the terminal solar wind velocity is unchanged (for all these cases, $u_E$ = 329 km/s). The altitudes of the maximum temperature are unchanged, and the values of T'$_{max}$ are only slightly affected; for all five cases $\sigma$ = 0.51, and $r_1$ = 1.31. Note that, at r = 1.31, the expansion velocity, u, is still small and subsonic.

Figure 7 illustrates the effect of changing the values of the parameters $r_1$ and $\sigma$, the altitude and width of the flux tube enhanced divergence rate. The values of these geometrical parameters are listed in Table 1 for the *Spv, SpvD, SpvDw1* and *SpvDh2w2* models; the corresponding values for T'$_{max}$, and h'$_{max}$ for the *dyn* and *shm* methods are shown in Table 2, for comparison.

It can be seen that the geometrical parameters characterizing the range of altitudes over which the divergence of the flux tube is concentrated can slightly influence the maximum temperature, its width and location.

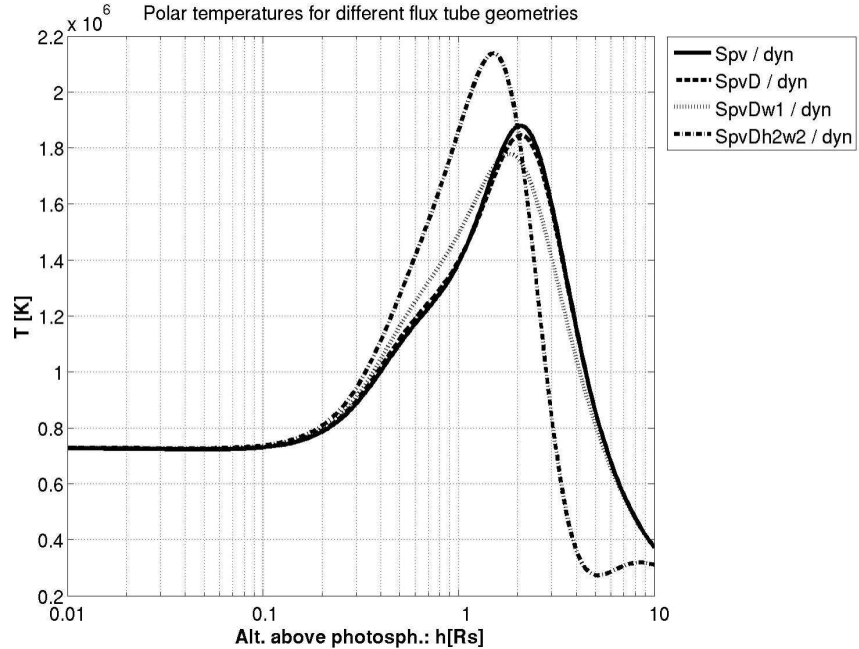

Figure 7. *Effect of the geometry of flux tubes.* All these *dyn* temperature distributions are derived from the same density profile: Saito's [27] average polar density model extended to match quiet solar wind density and bulk speed at 1AU ( $n_E$ = 2.22 cm$^{-3}$; $u_E$ = 329 km/s). The same maximum super-radial expansion factor ($f_{max}$ = 3) is assumed in all three cases, *SpvD, SpvDw1, SpvDh2w2*; only the two other geometrical parameters, $r_1$ and $\sigma$, characterizing the flow tubes, are different (see Table 1). It can be seen that the shapes of the temperature profiles and the temperature maxima (T'$_{max}$) are a more sensitive function of the flux tube expansion parameters $r_1$ and $\sigma$ than the value of $f_{max}$.

It should be noted that the results presented above are qualitatively different from those published by Kopp and Holzer [21], where the effect of changing $f_{max}$, was also modelled for coronal hole regions. Their approach is based on the standard numerical integration of the hydrodynamic transport equations in the Euler approximation/formulation; the working hypothesis of their model is that the temperature T(r) is proportional to n(r)$^\beta$, where $\beta$ is a constant polytropic index. Using standard hydrodynamic modelling algorithms, they determined the profile of u(r) along flux tubes whose cross-sections are given also by eq (11). From the calculated distribution of u(r), the profile of n(r) was derived by using the conservation mass flow. Figure 4 in [21] illustrates the family of hydrodynamic solutions that they obtained for u(r) using this approach. The critical solutions in their solar wind hydrodynamic models are characterized by an unusual maximum of u(r), which is located at low altitudes in the corona. Unlike the temperature profiles obtained here from eq. (10), their distribution of T(r) has no maximum in the inner corona, but continuously and smoothly decreases with r, in parallel with their solution for n(r) (see Figure 5 in [21]).

Kopp and Holzer's profiles for T(r) therefore differ greatly from the profiles obtained in this study. Here, u(r) is a smoothly increasing function of r (i.e., without a secondary peak at low coronal altitudes). It is actually T(r) that has a maximum in the inner corona. In other words, instead of the peak of coronal electron temperature obtained using the *dyn* method and illustrated in Figures 3-7, their polytropic temperature distribution is a smoothly decreasing function of r throughout the corona, up to 1AU and beyond. The reason for this important difference is that the calculated density distributions of Kopp and Holzer's polytropic model (given in their Figure 5) do not correspond closely enough to those inferred from the coronal WL brightness measurements shown in the Figure 1 above and used here to determine T(r).

Evidently, postulating an *ad hoc* polytropic relationship between T and n for the solar coronal plasma is not a realistic working hypothesis when there are distributed heat sources with a maximum energy deposition rate in the inner corona. Indeed, this leads to unexpected profiles for the coronal expansion bulk speed, with a peak of u(r) at low altitudes and fallible electron density distributions in the inner corona.

*Effects of $He^{++}$ ion concentration and proton temperature.* It has been assumed, so far, that the corona is fully ionized hydrogen plasma and that the proton temperature is equal to the electron temperature: $\tau_p = T_p/T_e = 1$. The presence of $He^{++}$ ions has been ignored. Let us now define their relative concentration by $\alpha = n_{He++}/n_p$, and assume their temperature $T_\alpha$ to be proportional to $T_e$ : $\tau_\alpha = T_\alpha / T_e$ . The baseline solutions presented earlier were based on the assumptions of $\tau_p = \tau_\alpha = 1$ and $\alpha = 0$. These assumptions are now relaxed, but the hypothesis that these additional free parameters are independent of r will be retained for convenience[19].

Since y(r), the dimensionless solutions of eq (15), are independent of $\alpha$, $\tau_p$ and $\tau_\alpha$, the temperature profiles, $T(r) = T^* y(r)$, for any other values of these free parameters can be easily generated from the baseline solutions calculated earlier for $\tau_p = \tau_\alpha = 1$ and $\alpha = 0$. Indeed, $T_e(r)$ is proportional to the '*intrinsic potential temperature*', $T^*$, which is defined by eq (8) or more generally by eq (A11) in Appendix A, where it is shown that $T^*$ is proportional to the factor:

$$\mu_e / \nu_p = (1 + 4\alpha) / (1 + 2\alpha + \tau_p + \alpha \tau_\alpha). \qquad (17)$$

For the definitions of $\mu_e$ and $\nu_p$, see eqs (A7) and (A8), respectively. For the baseline solutions, $\mu_e / \nu_p = ½$ and $T^* = 17$ MK. Other values for the '*intrinsic potential temperature*' $T^*$ can be determined from eq (A11) for other values of the input parameters. For $\alpha = 0.1$ and $\tau_p = \tau_\alpha = 1$, one obtains $T^* = 20.7$ MK. For other stars of mass, M', and radius, R', the following relationship holds: $T'^* = T^* M' R_S / M_S R'$.

It is interesting that the *hst* and *dyn* temperature profiles determined by $T(r) = T^* y(r)$ depends on the free parameters $\alpha$, $\tau_p$ and $\tau_\alpha$ only through the value of $T^*$. Unfortunately these free parameters cannot be inferred from WL brightness observations alone, but would have to be determined from other types of coronal measurements.

---

[19] Relaxing the latter assumption would make the determination of the coronal electron temperature distributions more difficult, and beyond our grasp, at least for the time being, because of the lack of available observations enabling a reliable determination of the values of $\alpha$, $\tau_p$ and $\tau_\alpha$ to be made and the lack of the variation in these parameters versus r.

**Conclusions**

The *shm* method commonly used to evaluate coronal temperatures from WL eclipse observations has been applied to a series of empirical coronal electron density distributions that are illustrated in Figure 1 as a function of h, the altitude above the photosphere. The coronal electron temperature distributions calculated by the popular *shm* method are shown in Figure 2, based on the assumption that hydrogen ions have the same temperature as the electrons, $\tau_p = T_p/T_e = 1$, and that $\alpha$, the concentration of alpha particles, is equal to zero.

This first set of calculations indicates that coronal temperature is not independent of r: a condition required, however, for the application of the *shm* method. This means that coronal temperature distributions derived using the *shm* method do not correspond to a corona in hydrostatic equilibrium.

The more adequate temperature distribution for which the corona is in hydrostatic equilibrium is determined by eq (10) corresponding to the asymptotic boundary condition: $T(r\rightarrow\infty) = 0$. This analytical solution was first reported by Alfvén [1] in 1941. In this widely known pioneering work, Alfvén introduced the concept of '*intrinsic potential temperature*', T*, which is defined by (8) and generalized by eq (A11) in Appendix A of the current paper.

This '*intrinsic potential temperature*' is a normalizing factor by which the dimensionless solution y(r) must be multiplied in order to obtain the coronal electron temperature that satisfies the hydrostatic eq (9) and that has the required regular asymptotic behaviour at infinity: $T_e(r\rightarrow\infty) = T^* \, y(r\rightarrow\infty) = 0$.

The electron temperature distributions were calculated using Alfvén's method for the different density profiles shown in Figure 1. The new solutions satisfying the hydrostatic equilibrium are labelled *hst*. The *hst* solution for Saito's extended equatorial density model is illustrated by the *Seq-hst* curve in Figure 3. The corresponding temperature profile *Seq-shm* obtained using the *shm* method is shown for comparison. This indicates that the maximum coronal temperature is located at a lower altitude when the *hst* method is used instead of the *shm* one. Both methods give almost the same value for the maximum temperatures, $T_{max}$. To some extent, this proves the reliability of eq. (10) and the *hst* method.

The *hst* temperature profile has a maximum value close to $r = 1.3\ R_S$. This is well above the top of the transition region separating the chromosphere and the corona: $r_{TR} = 1.0036\ R_S$. It is found that the maximum *hst* temperature is reached at higher altitudes over the polar regions or in coronal holes, rather than over the denser equatorial regions of the solar corona.

A third method was introduced to calculate the electron temperature distribution where and when the solar corona is expanding hydrodynamically, instead of being in hydrostatic equilibrium, as assumed for the *hst* method. This new method, *dyn*, provides the electron temperature $T_e(r)$ tending to zero at large distances, and satisfies Parker's hydrodynamic transport equations for the given empirical electron density profile.

The electron density ($n_E$) and bulk velocity ($u_E$) collected at 1AU from spacecraft observations were used to extend the eclipse empirical density profiles up to 1 AU. The third method for determining coronal temperatures from WL density profiles is outlined in Appendix A.

The *dyn* method was applied to the density profiles illustrated in Figure 1. These novel solutions were compared to the *hst* and *shm* solutions (see Figure 3). It was found that, in the equatorial region, the *dyn* and *hst* values were almost identical at $r < 3\ R_S$. Departures were found only at larger distances where the outward expansion velocity comes close to the sonic speed, and where the term corresponding to the solar wind acceleration in the hydrodynamic momentum eq (A6) becomes comparable to the kinetic pressure gradient.

Above polar regions and in coronal holes where coronal densities are smaller than over the equatorial region, the *dyn* temperature maximum is located at higher altitudes (see Figure 4). The calculated *dyn* temperature distributions are then rather sensitive to the value adopted for $u_E$, the solar wind speed at 1AU (see Figure 5). When $u_E$ is changed from zero to 600 km/s a the second peak of temperature, $T'_{max}$, emerges at $r \approx 3$, where u du/dr is maximum. For $u_E > 150$ km/s, this new temperature peak becomes larger than $T_{max}$, the one obtained using the *hst* method and corresponding to $u_E = 0$.

The effects of the geometrical flux/flow tube parameters $f_{max}$, $r_1$, $\sigma$ differ considerably from what is described by Kopp and Holzer [21] in their polytropic hydrodynamic solar wind model. Instead of obtaining unrealistic low altitude peaks for u(r) in the inner corona, as in the Kopp and Holzer's hydrodynamic models, the *dyn* method exhibits peaks of $T_e(r)$, the electron temperature profile in the inner corona. The latter occur where u du/dr, the rate of acceleration of the coronal plasma, is steepest and therefore where the coronal heating rate is expected to be at its maximum.

The ratio of proton and electron temperatures, $\tau_p$, and $\alpha$, the relative concentration of $He^{++}$ ions, are additional input parameters that influence both the value of T* and the overall electron temperature distribution, $T_e(r) = T^* y(r)$, where y(r) is the dimensionless solution given by eq (15).

Increasing $\tau_p$ and/or decreasing $\alpha$ reduces the electron temperature calculated using the *dyn* method. The altitudes of maximum temperatures ($h_{max}$ and $h'_{max}$), are given in Table 2 for the *dyn* and *shm* methods. The widths of the temperature peaks ($\Delta r$) are given only for the *dyn* method. This range of altitudes is related to the radial extent of the coronal heating source in the inner corona.

This study shows that ground-based and space measurements of coronal WL brightness could be exploited more extensively because they are expedient inputs for determining distributions of coronal temperatures with the highest spatial resolution. Obviously, the latter should be compared with other coronal temperatures inferred from other types of measurements.

The *dyn* method developed and applied in this study offers a unique diagnostic tool for determining the coronal temperature maximas, $T_{max}$ and $T'_{max}$, as well as the altitude where these peaks occur in the inner corona. The altitudes of temperature maximas are expected to correspond to those of maximum heat deposition rates. These outputs of the *dyn* method are therefore key diagnostic tools for investigating which physical mechanisms are mainly responsible for heating the solar corona, and thus resolving this long-standing issue in solar physics.

It would be interesting to apply the *dyn* method to the empirical static and dynamic tomographic 3-D electron density models of the solar corona recently developed by Butala et al. [39] based on STEREO observations. This could be used to determine corresponding 3-D distributions for coronal temperatures and to compare them with other independent measurements of temperatures in the solar corona.

Other more sophisticated techniques have been developed to evaluate coronal temperatures. They are based on radio-electric observations, as well as spectroscopic measurements in the UV,

EUV and Infra-red, and at Radio frequencies (see Golub and Pasachoff [35] and Aschwanden [5] for overviews).

It would be interesting to compare such complementary coronal temperature measurements with those obtained from WL eclipse observations in the inner corona determined by the *dyn* method described here.

From this study it can be seen that eclipse WL coronal observations still have great potential that has not yet been fully exploited. This observation reflects "the call for a concerted effort to support total eclipse observations over the next decade" recently proposed in a White Paper by Habbal et al. [36], albeit for other complementary scientific objectives.

**Appendix A:** Derivation of the differential equation whose solution gives the temperature distribution by the *dyn* method

Eqs (7) and (13) are derived from the stationary hydrodynamic equations for the transport of mass and momentum in the solar corona. Let us consider that the electrons and different ion species have the same average velocities, $u_j$. They are then all equal to the (single fluid) plasma bulk speed, $u$. Also, let us also postulate that the coronal and solar wind plasmas are homogeneous everywhere: $\alpha = n_{He^{++}}/n_p$ is then independent of r and time. Under these commonly adopted assumptions, the conservation of particle flux along flux tubes can be used to determine, from an empirical electron density profile, the distribution of the radial component of the bulk velocity:

$$u(r) = u_E [A_E / A(r)] [n_E / n_e(r)] \tag{A1}$$

A(r) is the cross-section flux/flow tube (given as an input as in [21]), and $n_e(r)$, the radial distribution of which is obtained from WL eclipse observations and *in situ* solar wind measurements at 1AU; the subscript E designates values at 1AU ($r = 215\ R_S$).

Following Kopp and Holzer [21], it became common practice to approximate A(r) by an exponential function of r:

$$A(r) / A(1) = r^2 f(r) = r^2 \{f_{max} \exp[(r - r_1)/\sigma] + f_1\} / \{\exp[(r - r_1)/\sigma] + 1\} \tag{A2}$$

where $f_1 = 1 + (1 - f_{max}) \exp[(1 - r_1)/\sigma]$; $f_{max}$ is the terminal super-radial expansion factor; $r_1$ is the heliocentric distance of largest expansion rate; and $\sigma$ is the range over which the super-radial expansion rate takes place. Note that when $f_{max} = 1$, then f(r) is independent of $r_1$ and $\sigma$; the radial expansion rate f(r) = 1 is then recovered, as in the early hydrodynamic and kinetic solar wind models. For $r > r_1 + 3\sigma$ is almost constant and equal to $f_{max}$.

When the average bulk velocities of the electrons and ions are all equal to u, the radial component of the steady state momentum transport equations of the electrons, protons and He$^{++}$ ions are:

$$n_e m_e u\, du/dr + dp_e/dr = -n_e m_e g - n_e e E \tag{A3}$$

$$n_p m_H u\, du/dr + dp_p/dr = -n_p m_H g + n_p e E \tag{A4}$$

$$4 n_\alpha m_H u\, du/dr + dp_\alpha/dr = -4 n_\alpha m_H g + 2 n_\alpha e E \tag{A5}$$

The gravitational acceleration is defined by $g(r) = G\, M_S / R_S^2\, r^2 = g_o / r^2$.

The concatenation of the eqs (A3), (A4) and A(5) leads to the familiar (single fluid) hydrodynamic momentum equation:

$$\rho\, u\, du/dr + dp/dr = \rho\, g \qquad (A6)$$

Note that the net electrostatic force density ($\Sigma_i\, Z_i\, n_j\, e\, E$) has disappeared from (A6) because of the quasi-neutral condition: $n_e = n_p + 2\, n_\alpha$, or $n_e = n_p (1 + 2\, \alpha)$ where $\alpha = n_\alpha/n_p$.

The mass density of the plasma is equal to $\rho = n_p\, m_H + 4\, n_\alpha\, m_H + n_e\, m_e \approx (n_p + 4\, n_\alpha)\, m_H = n_p (1 + 4\, \alpha)\, m_H = n_e\, \mu_e\, m_H$ where:

$$\mu_e = (1 + 4\, \alpha) / (1 + 2\, \alpha) \qquad (A7)$$

In eq (A6), the total kinetic pressure is $p = p_e + p_p + p_\alpha = \nu_p\, n_e\, k\, T_e$ where:

$$\nu_p = (1 + 2\, \alpha + \tau_p + \alpha\, \tau_\alpha) / (1 + 2\, \alpha) \qquad (A8)$$

where the ratios between the proton or alpha temperature and electron temperature are defined by $\tau_{p(\alpha)} = T_{p(\alpha)}/T_e$.

When the temperature ratios $\tau_p$ and $\tau_\alpha$ are constants independent of r, the constants $\mu_e$ and $\nu_p$ are also independent of r.

When $\alpha$ is independent of r, the coronal plasma is homogeneous. These additional hypotheses are used here as simplifying assumptions that could be put aside in future modelling efforts. At this stage, these assumptions serve to outline the effects of some input parameters on the calculated *dyn* temperature distributions. This implies that the results presented here should be viewed as exploratory steps that might lead to future ones where and when some of the simplifying assumptions can be put aside when more comprehensive experimental measurements become available.

Note that as a result of the small size of the electron mass compared with ion masses, (A3) can be approximated by:

$$k\, T_e\, d\ln n_e/dr + d(k\, T_e)/dr \approx -\, e\, E \qquad (A9)$$

This equation enables the polarization electric field distribution, $E(r)$, to be evaluated when the empirical distribution for $n_e(r)$ is known (from observations) and when the electron temperature distribution, $T_e(r)$, has been calculated for this same electron density distribution using the *dyn* method described in this paper.

Replacing $\rho$ and $p$ by the expressions given above, and assuming that $\mu_e$ and $\nu_p$ are independent of r, (A6) becomes:

$$n_e\, \mu_e\, m_H\, u\, du/dr + \nu_p\, k\, T_e\, dn_e/dr + \nu_p\, n_e\, k\, dT_e/dr = -\, n_e\, m_H\, \mu_e\, g_o/r^2 \qquad (A10)$$

Let us redefine Alfvén's *'intrinsic potential temperature'* (8) by:

$$T^* = G M_S m_H \mu_e / \nu_p k R_S = g_o R_S m_H \mu_e / \nu_p k \tag{A11}$$

When $T_p = T_\alpha = T_e$ and $\alpha = 0$, it can be verified that $\mu_e / \nu_p = \frac{1}{2}$ and $T^* = 17$ MK. The definition (A11) is more general than (8), and should be used when the concentration of $He^{++}$ ions has to be taken into account and when the temperatures of the ions are not equal to those of the electrons.

Using the solar radius ($R_S$) as a unit for r, and $y = T_e / T^*$, eq (A11) becomes:

$$dy/dr + y \, d \ln n_e/dr = -(1/r^2) + (m_H \mu_e / \nu_p k T^*) u \, du/dr \tag{A12}$$

The lhs corresponds hydrostatic pressure force. The first term in the rhs corresponds to the downward gravitational acceleration, and the second term is the inertia force opposed to the outward acceleration of coronal plasma.

It can be verified that the analytical solution of this equation for which $y(r) = 0$ when $r \to \infty$ is simply given by:

$$y = [1/n_e(r)] \int_\infty^r \{n_e(r')/r'^2\} [1 + F(r')] \, dr' \tag{A13}$$

where
$$F(r) = (m_H \mu_e / \nu_p k T^*) r^2 u \, du/dr = r^2 (u \, du/dr) / g_o R_S \tag{A14}$$

$F(r)$ is the ratio of the radial components of the hydrodynamic acceleration and the gravitational deceleration; it is an analytical function of r that is fully determined by eqs (A1) & (A2), since the distribution of $n_e(r)$ is given here as an empirical function of r. It can be verified that $\lim F(r \to \infty) = 0$.

Note that eq A(13) is a generalization of the hydrostatic solution reported by Alfvén (1941). When $u(r) \equiv 0$, then $F(r) \equiv 0$ and eq (A13) become identical to eq (9). The new *dyn* solution (A13) - (A14) is more general because it enables the electron temperature profiles to be calculated when the solar corona expands hydrodynamically, rather than being in hydrostatic equilibrium, as thought until 1958 [10] [25].

**Acknowledgements**
The author thanks Clément Botquin and Koen Stegen for their efficiency in building and maintaining the computer algorithms. He also wishes to thank the Belgian Institute for Space Aeronomy (BISA) for the financial and logistic support. He much appreciated the valuable comments and constructive suggestions from Marius Echim (BISA), Serge Koutchmy (IAP), Hervé Lamy (BISA), Viviane Pierrard (BISA), Cyril Simon Wedlund (BISA), Mea Simon Wedlund (Univ. of Otago), and Yuriy Voitenko (BISA).

## Bibliographical references